\documentclass[prb,showpacs,twocolumn,aps,superscriptaddress,a4paper]{revtex4-1}
\usepackage{dcolumn,amssymb,amsmath,amsfonts,graphicx,latexsym,color}
\usepackage{epstopdf}
\begin{document}

\title{Localization and mobility edges in the off-diagonal quasiperiodic model with slowly varying potentials}
\author{Tong Liu}
\affiliation{Department of Physics, Southeast University, Nanjing 211189, China}
\author{Gao Xianlong}
\affiliation{Department of Physics, Zhejiang Normal University, Jinhua 321004, China}
\author{Shihua Chen}
\affiliation{Department of Physics, Southeast University, Nanjing 211189, China}
\author{Hao Guo}
\thanks{Corresponding author: guohao.ph@seu.edu.cn}
\affiliation{Department of Physics, Southeast University, Nanjing 211189, China}

\begin{abstract}
We study a one-dimensional system that includes both a commensurate off-diagonal modulation of the hopping amplitude and an incommensurate, slowly varying diagonal on-site modulation. By using asymptotic heuristic arguments, we identify four closed form expressions for the mobility edges. We further study numerically the inverse participation ratio, the density of states and the Lyapunov exponent. The numerical results are in exact agreement with our theoretical predictions. Besides a metal-insulator transition driven by the strength of the slowly varying potential, another four insulator-metal
transitions are found in this model as the energy is increased in magnitude
from the band center ($E =0$) to the mobility edges ($\pm E_{c2}, \pm E_{c1}$).
\end{abstract}

\maketitle

\section{Introduction}
\label{n1}

Anderson localization~\cite{1an}, the absence of diffusion of matter waves in disorder systems, is an active research subject in condensed matter physics due to its importance and complexities. With regard to the randomly-distributed potentials the scaling theory~\cite{2scal} predicts
that there is no metal-insulator transition in one-dimensional (1D) systems. Hence all wave-functions are exponentially localized no matter how small the existed disorder is. However, the emergence of  quasiperiodic/incommensurate potentials in 1D systems~\cite{2PRA,mod,3PRA,He,qian,Gramsch,14PRL,15PRL,16PRB,17PRL,LiuG} goes beyond this prediction.
As an important paradigm, the Aubry-Andr\'{e} (AA) model~\cite{7aubry} can undergo a transition from the extended state to the localized state as the amplitude of the incommensurate potential increases. The nature of the AA model has been well understood with extensive researches~\cite{21SO,22SOU,dei,lah,cao,Mo1}. It is well known that at the phase transition point the spectrum of the AA model is a Cantor set and all wave-functions are critical, i.e., neither extended nor localized. However, unlike the Anderson transition in 3D situations, there is no mobility edge~\cite{D26,J27,S28} in the AA model, where all wave-functions are either extended or localized (critical), depending on the relative strength of the incommensurate potential and the hopping amplitude.

The concept of the mobility edge, first proposed by Mott~\cite{mott} regarding the 3D Anderson model, is signaled by an energy-level border separating the localized and extended bands. Mott argued, if the extended and localized states can coexist at the
same spectral energy level, then, an arbitrarily small energy perturbation leads to the hybrid between localized and extended states. As a result, only extended states can remain, which is not true. However, whether a 1D system can host mobility edges is another interesting question. About thirty years ago a unique class of systems with 1D potentials~\cite{hir,gri,thou,sarma,tong}, i.e., very slowly varying incommensurate potentials in real space, was introduced and extensively studied. These deterministic (diagonal) potentials are neither random nor simply incommensurate. The important finding by Sarma et. al.~\cite{sarma} is that, there is a metal-insulator transition in this model with the mobility edges located at certain energies with the eigenstates at the band center being all extended whereas the band-edge states all localized.
Since then, different variations of the 1D models containing mobility edges have been studied, including those with a rational term added to the slowly varying incommensurate potential~\cite{tong} or the Aubry-Andr\'{e}-like model with a duality symmetry~\cite{12PRL,13PRL}. Recently, the off-diagonal AA model, an extension of the AA model including both commensurate and incommensurate off-diagonal
hopping modulations, has attracted much attention due to the arising rich and novel quantum phases, especially the important nontrivial zero-energy topological edge states~\cite{18PRB,19PRB,liu}.

Here we include an off-diagonal commensurate modulation in Sarma's model with slowly varying incommensurate diagonal modulations. While the latter brings out a metal-insulator transition, we find that, the former modulate the mobility edges greatly. This result opens perspectives for further investigations on the interplay between mobility edge and off-diagonal modulations.

Our generalized AA model, including commensurate off-diagonal and slowly varying incommensurate diagonal modulations, is expressed as,
\begin{equation}\label{tb1}
    \hat H=-\sum_{i=1}^{L-1}(t+\lambda_{i})(\hat{c}_{i}^\dag \hat{c}_{i+1}+h.c.)+\sum_{i=1}^{L}V_{i} \hat{n}_{i},
\end{equation}
where $\hat{c}_{i}$ is the fermionic annihilation operator, $\hat{n}_i=\hat{c}^\dagger_{i}\hat{c}_{i}$ is the particle number operator,
$L$ is the total number of sites, $\lambda_{i}=\lambda\cos(2\pi b{i})$ with $0<\lambda<1$ being the strength of the commensurate modulation on the off-diagonal hopping amplitude, and $V_{i}=V\cos(2\pi\beta{i^{v}}+\phi)$ with $0<v<1$ and $V>0$ being the strength of the slowly varying incommensurate potential. Without loss of generality, we choose the parameters $b=1/2$, $\beta=(\sqrt{5}-1)/2$, the phase in the incommensurate modulation $\phi = 0$ and $v = 0.5$. For convenience, $t = 1$ is set as the energy unit.

When $\lambda = 0$ and $v = 1$, our model reduces to the AA model, while $\lambda = 0$ and $0<v<1$ is the extended AA model studied by Sarma et. al.~\cite{sarma}. It is well known that this model has two mobility edges located at energies $\pm |2-V|$. When $V<2$, all wave-functions with eigenenergies in $[V-2,2-V]$ are extended and otherwise localized. When $V>2$, all wave-functions are localized, which is same as that in the AA model.
In this paper, we focus on the situation with $0<\lambda<1$ and $0<v<1$.
The main findings are, (i) when $V<1-\lambda$ there exist two additional mobility edges located at $\pm | 2\lambda+V |$ besides those at $\pm|2-V|$, (ii) the all-wave-function-localized transition point is located at $V=1-\lambda$.

The rest of the paper is organized as follows. In Sec.~\ref{n2}, we theoretically
give the heuristic arguments for the Hamiltonian~(\ref{tb1}).
In Sec.~\ref{n3}, we present our numerical results and compare them with the theoretical analysis. The conclusion is summarized in Sec.~\ref{n4}.

\section{HEURISTIC ARGUMENTS}
\label{n2}
The slowly varying incommensurate potential brings up new nature of localization, which is substantially different from the 3D Anderson model~\cite{sarma}. To demonstrate the existence of mobility edges, Sarma et. al.~\cite{sarma} presented some preliminary heuristic arguments and applied the semiclassical WKB technique. Here we also implement this method to obtain the explicit expressions of mobility edges.

By noticing that the slowly varying potential difference of $V_{i}=V\cos(2\pi\beta{i^{v}})$
vanishes in the thermodynamic limit~\cite{sarma}, we thus write
\begin{equation}\label{tb2}
\frac{dV_{i}}{di}=-2V\pi\beta i^{v-1}\sin(2\pi\beta{i^{v}}).
\end{equation}

When $ i \rightarrow \infty$, Eq.~(\ref{tb2}) can be written as
\begin{equation}\label{tb3}
\lim_{i\rightarrow\infty} \left\vert\frac{dV_{i}}{di}\right\vert=-\lim_{i\rightarrow\infty}  2V\pi\beta \frac{|\sin(2\pi\beta{i^{v}})|}{i^{1-v}}=0,
\end{equation}
since $0<v<1$. Equivalently, the potential difference $ V_{i+1} - V_{i}\rightarrow 0$ when the lattice number $i$ is large enough, i.e., the potential $V_{i}$ becomes a constant. This asymptotic property of ``being constant" of $V_{i}$ is crucial for the localization
property of this model. On the other hand, another unique characteristic of our model is the existence of the off-diagonal commensurate modulation $\lambda_{i}$ which leads to a parity symmetry. The Schr\"{o}dinger equations for Eq.~(\ref{tb1}) rewritten according to the odd and the even lattice site become,
\begin{equation}\label{tb4}
\begin{split}
    & (1-\lambda) \psi_{2m} + (1+\lambda) \psi_{2m-2} + C\psi_{2m-1}= 0,\\
    & (1+\lambda) \psi_{2m+1} + (1-\lambda) \psi_{2m-1} + C\psi_{2m}= 0,\\
    & (1-\lambda) \psi_{2m+2} + (1+\lambda) \psi_{2m} + C\psi_{2m+1}= 0,
\end{split}
\end{equation}
where $m$ is an arbitrary positive integer and $C=E-V\cos(2\pi\beta{(2m-1)^{v}})=E-V\cos(2\pi\beta{(2m)^{v}})=E-V\cos(2\pi\beta{(2m+1)^{v}})$ since  all the $V_{i}$ are constant. From Eq.~(\ref{tb4}) we obtain
\begin{equation}
\psi_{2m+2} + \frac{2+2 \lambda^2-C^2}{1-\lambda^2}\psi_{2m}+ \psi_{2m-2}= 0.
\label{tb5}
\end{equation}
Following the asymptotic heuristic argument~\cite{sarma}, we write $\psi_{2m}\sim Z^{m}$, where $Z$ is a complex quantity. Then Eq.~(\ref{tb5}) becomes
\begin{equation}
Z^2 + \frac{2+2 \lambda^2-C^2}{1-\lambda^2}Z+ 1= 0,
\label{tb6}
\end{equation}
to which the complex solutions are
\begin{equation}
Z_{1,2}=\frac{-G\pm\sqrt{G^2-4}}{2}
\label{tb7}
\end{equation}
 with $G=\frac{2+2 \lambda^2-C^2}{1-\lambda^2}$. From Eq.~(\ref{tb7}) we conclude that the amplitude is complex or extended (due to $|Z|=1$) if $| G|<2$ whereas real or localized if $| G|>2$. Note $C_\textrm{max}= | E| +V$ and $C_\textrm{min}=| E| -V$, so if $E$ and $V$ are fixed, we get
\begin{equation}
\begin{split}
    & G_\textrm{max} = \frac{2+2 \lambda^2-(| E| -V)^2}{1-\lambda^2},\\
    & G_\textrm{min} = \frac{2+2 \lambda^2-(| E| +V)^2}{1-\lambda^2}.
    \label{tb8}
\end{split}
\end{equation}
The conditions for extended and localized solutions are respectively given by
\begin{equation}
\begin{split}
    & G_\textrm{max}<2 \Rightarrow 2\lambda + V<| E| (\textrm{extended}),\\
    & G_\textrm{max}>2 \Rightarrow 2\lambda + V>| E| (\textrm{localized}),\\
    & G_\textrm{min}>-2 \Rightarrow 2- V>| E| (\textrm{extended}),\\
    & G_\textrm{min}<-2 \Rightarrow 2- V<| E| (\textrm{localized}).
    \label{tb9}
\end{split}
\end{equation}
Note that for the existence of the mobility edges there is an implicit condition that $2-V>2\lambda + V$, i.e., $V<1-\lambda$. If this condition is satisfied there will be four mobility edges with $\pm E_{c1}=\pm \vert 2- V\vert$ and $\pm E_{c2}=\pm \vert 2\lambda + V\vert$.
That is, the model defined by the Hamiltonian~(\ref{tb1}) with
$0<v<1$ and $V<1-\lambda$ has localized states at the band center ($-E_{c2}<E<E_{c2}$) and at the band edges ($E>E_{c1}$, $E<-E_{c1}$), extended states at $-E_{c1}<E<-E_{c2}$ and $E_{c2}<E<E_{c1}$.

Eq.~(\ref{tb9}) also implies that there are four insulator-metal
transitions in the system as the energy is increased in magnitude
from the band center ($E =0$) to the band edges
($\pm E_{c2}, \pm E_{c1}$).

When $V>1-\lambda$, all the states are localized and there is no extended state, which means that $V=1-\lambda$ is a
transition point from a state of mobility edge to a state of all the wave-functions localized.

\section{NUMERICAL RESULTS}
\label{n3}

In this section we present detailed numerical analysis to support the theoretical predications given in the previous section, which is exact in the thermodynamic limit. We directly diagonalize the model Hamiltonian (\ref{tb1}) to obtain the eigenenergies $E$ and the corresponding wave-functions $\psi$. After that, it is fairly straightforward to calculate the typical physical quantities used in the disordered system, such as the inverse participation ratio and the Lyapunov exponent, to distinguish the localized and extended states. All the numerical results are in exact agreement with the theoretical predictions in Sec.~\ref{n2}, demonstrating that the asymptotic theory works very well even for the finite-sized systems of which the potential is not strictly a constant.
\begin{figure}
  \centering
  \includegraphics[width=0.5\textwidth]{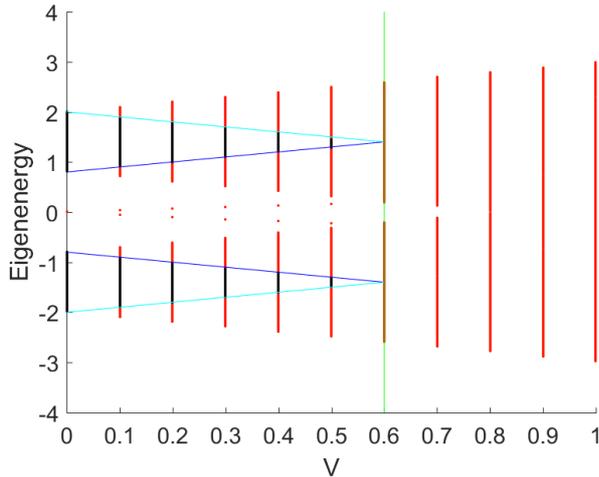}\\
  \caption{(Color online) Eigenenergy of Eq.~(\ref{tb1}) as a function of $V$ with $\lambda=0.4$. The total number of sites is set to be $L=10000$ hereafter in this paper. The magnitude of the IPR of the extended state accordingly approaches $10^{-4}$. Different colours of the eigenenergy curves indicate different magnitudes of the IPR of the corresponding wave-functions. The black eigenenergy curves denote the extended states (the magnitude of the IPR $< 10^{-3}$), and the red eigenenergy curves denote the localized states (the magnitude of the IPR $> 10^{-3}$). The cyan solid lines represent two mobility edges with $E_{c1}=\pm (2- V)$, and the blue solid lines represent another two mobility edges with $E_{c2}=\pm (2\lambda + V)=\pm (0.8 + V)$. It is clear that these four boundaries are exactly located between spatially localized and extended states. The green solid line represents the all-wave-function-localized phase transition point $V=1-\lambda=0.6$.}
  \label{001}
\end{figure}

First we calculate the inverse participation ratio (IPR). The IPR of a normalized wave function $\psi$ is defined as~\cite{IPR1,IPR2,IPR3},
\begin{equation}
\text{IPR}_n =\sum_{j=1}^{L} \left|\psi^n_{j}\right|^{4},
\end{equation}
where $L$ denotes the total number of sites and $n$ is the energy level index.
It is well known that the IPR of the extended state scales like $L^{-1}$, which approaches $0$ in the thermodynamic limit, but finite for a localized state.

Figure~\ref{001} plots the eigenenergy and IPR of the corresponding wave-functions of the Hamiltonian (\ref{tb1}) as a function of the incommensurate potential strength $V$. The blue and cyan solid lines represent mobility edges given by Eq.~(\ref{tb9}). As expected from the asymptotic heuristic argument, when $V<1-\lambda=0.6$
there are four metal-insulator transitions in the spectrum
(the magnitude of the IPR $> 10^{-3}$ or $< 10^{-3}$). The corresponding wave-functions vary from above to bottom according to `localized$\rightarrow$extended$\rightarrow$localized$\rightarrow$extended$\rightarrow$localized', indicating that Eq.~(\ref{tb9}) indeed defines four mobility edges at $\pm E_{c1}$ and $\pm E_{c2}$. When $V>1-\lambda=0.6$ all red eigenenergy curves indicate all the states are localized and there is no extended state, which also agrees with the asymptotic prediction given in Sec.~\ref{n2}.
As a result, a metal-insulator transition happens at $V=1-\lambda=0.6$. Another four insulator-metal
transitions are found in this model as the energy is increased in magnitude
from the band center ($E =0$) to the mobility edges ($\pm E_{c2}, \pm E_{c1}$).

\begin{figure}
  \centering
  \includegraphics[width=0.5\textwidth]{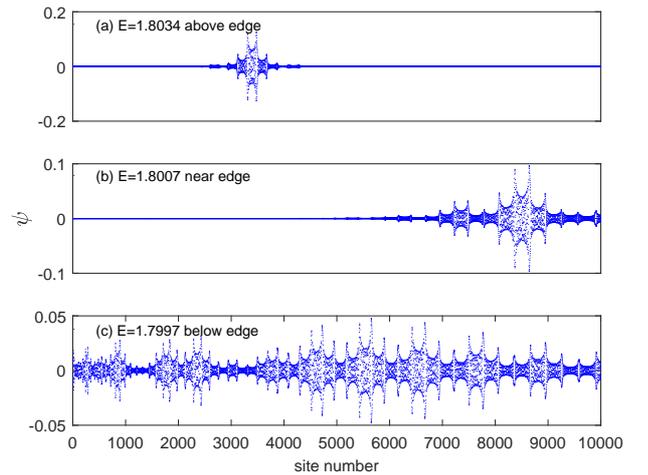}\\
  \caption{(Color online) Eigenstates of Eq.~(\ref{tb1}) with $\lambda=0.4$ and $V=0.2$ around the mobility edge with $E_{c1}=2- V=1.8$. Three different eigenenergies: (a) high energy localized state above the mobility edge, (b) critical state near the mobility edge, and (c) low energy extended state below the mobility edge.}
  \label{002}
\end{figure}

\begin{figure}
  \centering
  \includegraphics[width=0.5\textwidth]{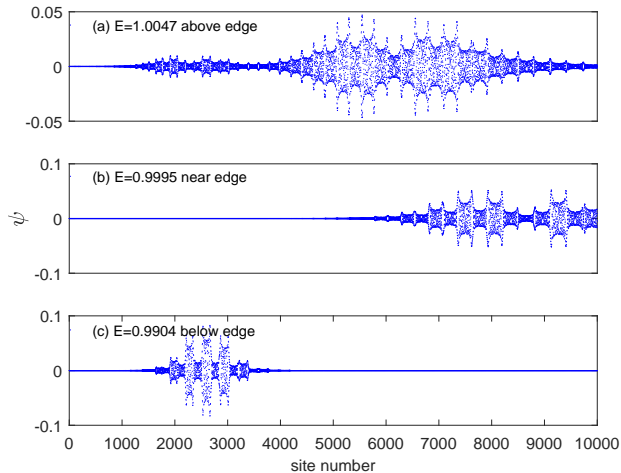}\\
  \caption{(Color online) Eigenstates of Eq.~(\ref{tb1}) with $\lambda=0.4$ and $V=0.2$ around the mobility edge with $E_{c2}=2\lambda + V=1$. Three different eigenenergies: (a) high energy extended state above the mobility edge, (b) critical state near the mobility edge, and (c) low energy localized state below the mobility edge.}
  \label{003}
\end{figure}

Fig.~\ref{002} and Fig.~\ref{003} show eigenstates  at $\lambda=0.4$ and $V=0.2$ for three different eigenenergy levels, which are located above, near, and below two mobility edges with $E_{c1}=2- V=1.8$ and $E_{c2}=2\lambda + V=1$, respectively. From Fig.~\ref{002} we can see that at the mobility edge with $E_{c1}=2- V=1.8$ the wave-function is localized at high energy levels [Fig.~\ref{002}(a)], critical near
the boundary [Fig.~\ref{002}(b)] and extended at low energy levels [Fig.~\ref{002}(c)]. In contrast, Fig.~\ref{003} shows that at the mobility edge with $E_{c2}=2\lambda + V=1$ the wave-function is extended at high energy levels [Fig.~\ref{003}(a)], critical near the boundary [Fig.~\ref{003}(b)] and localized at low energy levels [Fig.~\ref{003}(c)].

To strengthen our findings, we calculate the density of states $D(E)$ and the Lyapunov exponent $\gamma(E)$ of this quaisiperiodic system. Here $D(E)$ is defined as,
\begin{equation}
D(E) =\sum_{n=1}^{L} \delta(E-E_n),
\end{equation}
and $\gamma(E)$~\cite{sarma,tong} is,
\begin{equation}
\gamma(E_n) =\frac{1}{L-1}\sum_{n\neq m}^{L}\ln| E_n-E_m|,
\end{equation}
where $E_n$ is the $n$-th eigenenergy and $L$ is the total number of sites. The Lyapunov exponent is the inverse localization length, hence $\gamma=0$ for an extended state whereas $\gamma \neq0$ for a localized state. The density of states and the Lyapunov exponent are connected by the relation
\begin{equation}
\gamma(E) =\int dE'D(E') \ln| E- E'|.
\label{tb13}
\end{equation}

\begin{figure}
  \centering
  \includegraphics[width=0.5\textwidth]{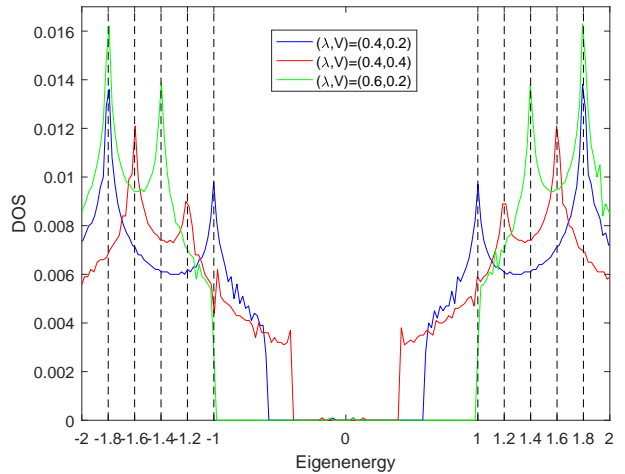}\\
  \caption{(Color online) Density of states (DOS) of Eq.~(\ref{tb1}) as a function of eigenenergy for parameters $(\lambda,V)=(0.4,0.2)$, $(0.4,0.4)$ and $(0.6,0.2)$. There is clearly a dramatic change when the eigenenergy passes through the mobility edges with $E_{c1}=\pm (2- V)$ and $E_{c2}=\pm (2\lambda + V)$.  }
  \label{004}
\end{figure}

\begin{figure}
  \centering
  \includegraphics[width=0.5\textwidth]{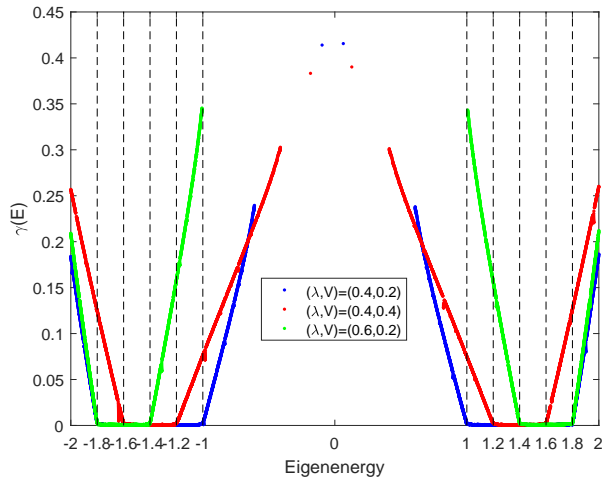}\\
  \caption{(Color online) The Lyapunov exponent $\gamma(E)$ vs eigenenergy for parameter sets $(\lambda,V)=(0.4,0.2)$, $(0.4,0.4)$ and $(0.6,0.2)$. If the eigenenergy is in either one of the intervals $[V-2, -2\lambda - V]$ and $[2\lambda + V, 2 - V]$, $\gamma(E)$ approaches $0$, indicating the corresponding extended state. Whereas if the eigenenergy is outside both intervals, $\gamma(E)$ is finite, indicating the corresponding localized state. }
  \label{005}
\end{figure}

In Fig.~\ref{004} we show our numerical results for the density of states $D(E)$. In order to ensure the universality of the results we choose three different sets of parameters $(\lambda,V)=(0.4,0.2)$, $(0.4,0.4)$ and $(0.6,0.2)$. One can see that the energy band consists of two subbands which are quasi-symmetric about $E=0$. The singularity of the density of states at the spectrum, at which the transition occurs~\cite{Thoul}, reflects the change of the nature of the eigenstates. Therefore in Fig.~\ref{004} two sharp peaks in both subbands indicate the extended-localized transitions corresponding to two mobility edges located at $| E_{c1}|=2- V$ and $| E_{c2}|=2\lambda + V$. In Fig.~\ref{005} we plot the Lyapunov exponent $\gamma(E)$ for the three same sets of parameters as previous. All numerical results for $\gamma(E)$ are in excellent agreement with the theoretical predictions just as those for the density of states. We have also checked other combinations of parameters and get the same results as expected.

Another interesting subject is the critical exponents of the Lyapunov exponent at the mobility edge defined by
\begin{equation}
\gamma(E) \sim | E- E_{c i}|^\theta,\, i=1,2,\,
\end{equation}
in the localized regions of energy spectrum. Similarly, the behavior of the density of states at the mobility edge can be written as
\begin{equation}
D(E) \sim | E- E_{c i}|^{-\delta},\, i=1,2.
\end{equation}
Therefore, according to Eq.~(\ref{tb13}) the critical exponents $\theta$ and $\delta$ are clearly related by the equation
\begin{equation}
\theta + \delta =1.
\end{equation}

From Fig.~\ref{005}, we notice that the singularities of $\gamma(E)$ at both mobility edges with $E_{c1}=\pm (2- V)$ and $E_{c2}=\pm (2\lambda + V)$ are identified to be linear indicating the critical exponent $\theta=1$, same as that in the previous model~\cite{sarma}. Unlike the models with higher-dimensional randomly-distributed potentials, the density of states in this model is not smooth through the mobility edge, but logarithmically divergent at the mobility edge just as the Sarma's model~\cite{sarma}. Therefore the critical exponent of the density of states is $\delta=0$. 
We have checked that the parameters $\lambda$, $V$, $\beta$ and $v$ are all irrelevant with regard to the critical exponents $\theta$ and $\delta$.

It is clear from our numerical results that the asymptotic theory developed in Sec.~\ref{n2} is essentially exact for
describing the localization properties of the off-diagonal quasiperiodic model with slowly varying potentials (for $0<v<1$). In addition, we find that the mobility edges are dependent on the parameters $t$, $\lambda$ and $V$ but irrelevant to $b$, $\beta$ and $v$.

\section{Conclusions}
\label{n4}

In summary, we have studied the localization properties of a class of off-diagonal quasiperiodic models with a slowly varying incommensurate on-site potential. In the thermodynamic limit this slowly varying incommensurate potential becomes a constant. Making use of this asymptotic heuristic argument, we
reveal that there exist four mobility edges separating localized and extended states when the incommensurate potential strength $V$ is smaller than certain critical values. When $V$ goes above this threshold all wave-functions are localized. By diagonalizing the discretized Hamiltonian, we numerically
solve the Schr\"{o}dinger equation. Our results clearly show the
existence of mobility edges and are in excellent agreement with the theoretical predictions for the mobility edges by analysing the inverse participation ratio, the local density of state, and the Lyapunov exponent.

We believe that the interesting features of this model will bring a new perspective to a wide range of quasiperiodic systems, which is of great interest in the present cold atomic experiments~\cite{Mo2,Sc1,roati}.

\begin{acknowledgments}
G. H. thanks the support from the NSF of China (Grant No.
11674051).
G.X. was supported by the NSF of Zhejiang Province (Grant No. Z15A050001),
the NSF of China (Grant No. 11374266), and the Program for New Century Excellent Talents in University.

\end{acknowledgments}

\end{document}